\documentclass[twocolumn,final]{elsarticle}
\usepackage{graphicx}
\usepackage{braket}

\journal{Journal of Quantitative Spectroscopy and Radiative Transfer}

\usepackage{graphicx}
\usepackage{amsmath}
\usepackage{amssymb}

\def\sc {\scriptscriptstyle}

\begin{document}

\begin{frontmatter}

\title{Rayleigh-Brillouin light scattering spectroscopy of nitrous oxide (N$_2$O)}

\author[vua]{Y. Wang}
\author[hus]{K. Liang}
\author[tue]{W. van de Water}
\author[cur]{W. Marques Jr.}
\author[vua]{W. Ubachs\corref{mycorrespondingauthor}}\cortext[mycorrespondingauthor]{Corresponding author}\ead{w.m.g.ubachs@vu.nl}

\address[vua]{Department of Physics and Astronomy, LaserLaB, Vrije Universiteit, De Boelelaan 1081, 1081 HV Amsterdam, The Netherlands}
\address[hus]{School of Electronic Information and Communications, Huazhong University of Science and Technology, Wuhan 430074, China}
\address[tue]{Physics Department, Eindhoven University of Technology, Postbus 513, 5600 MB Eindhoven, The Netherlands}
\address[cur]{Departamento de F\'{i}sica, Universidade Federal do Paran\'{a}, Caixa Postal 10944, 81531-990 Curitiba, Brazil}

\begin{abstract}
\begin{small}
\noindent
High signal-to-noise and high-resolution light scattering spectra are measured for nitrous oxide (N$_2$O) gas at an incident wavelength of 403.00 nm, at 90$^\circ$ scattering, at room temperature and at gas pressures in the range $0.5-4$ bar. The resulting Rayleigh-Brillouin light scattering spectra are compared to a number of models describing in an approximate manner the collisional dynamics and energy transfer in this gaseous medium of this polyatomic molecular species. The Tenti-S6 model, based on macroscopic gas transport coefficients, reproduces the scattering profiles in the entire pressure range at less than 2\% deviation at a similar level as does the alternative kinetic Grad's 6-moment model, which is based on the internal collisional relaxation as a decisive parameter. A hydrodynamic model fails to reproduce experimental spectra for the low pressures of 0.5-1 bar, but yields very good agreement ($< 1$\%) in the pressure range $2-4$ bar. While these three models have a different physical basis the internal molecular relaxation derived can for all three be described in terms of a bulk viscosity of $\eta_b \sim (6 \pm 2) \times 10^{-5}$ Pa$\cdot$s.
A 'rough-sphere' model, previously shown to be effective to describe light scattering in SF$_6$ gas, is not found to be suitable, likely in view of the non-sphericity and asymmetry of the N-N-O structured linear polyatomic molecule.
\end{small}
\end{abstract}


\end{frontmatter}


\section{Introduction}
\label{SecN2OblueIntroduction}

Spontaneous Rayleigh and Brillouin scattering arises from fluctuations in the dielectric constant of gases and its spectral profiles have been studied since the early 20$^{th}$ century~\cite{Strutt1899,Brillouin1922,Mandelstam1926}. The density fluctuations associated with molecular thermal motion takes the form of entropy fluctuations at constant pressure causing a central elastic Rayleigh scattering peak. Pressure fluctuations in the form of acoustic waves, at constant entropy, cause inelastic side-peaks referred to as Brillouin scattering~\cite{Fabelinskii2012}. In the decade after the invention of the laser numerous studies have been performed measuring the spectral profile of Rayleigh-Brillouin scattering with the goal to derive collisional properties of gaseous media~\cite{Greytak1966,Lao1976a,Sandoval1976,Letamendia1982}. More recently the field of RB-scattering has been revived with the goal to monitor gas flow and the conditions of the Earth's atmosphere~\cite{Lock1992,Witschas2014,Witschas2014b,Gu2014b}. Independently, in addition to conventional spontaneous RB scattering, methods of coherent RB scattering have been developed for the investigation of collisional phenomena in gases \cite{Pan2002,Pan2004,Bookey2007,Meijer2010}.

Typically, at different gas densities, the spectral lineshape of the scattered light intensity will be different, thereby reflecting the collisional phenomena occurring in the gas. A key scaling parameter is the uniformity parameter $y$, which compares the reciprocal of the scattering wave vector $k_{\rm{sc}}$ to the mean free path between collisions $l_{\rm{mfp}}$, hence $y = 1/k_{\rm{sc}}l_{\rm{mfp}}$. When $y$ becomes large with respect to unity, such as for dense gases, the effect of  acoustic modes will become apparent as side peaks in the spectra profile. The frequency shift of these Brillouin side features is $f_s = \upsilon_{s}k_{\rm{sc}}/2\pi$, with $\upsilon_{s}$ the speed of a sound wave in the dense gas. The density fluctuations in this hydrodynamic regime can be described by the Navier-Stokes equation. The broadening effects are homogeneous and the central and both Stokes peaks adopt a Lorentzian functional form. In the opposite case, of the Knudsen regime for values $y \ll 1$, the spectral line shape adopts the character of a pure Gaussian, as a result of the inhomogeneous effect of molecular random thermal motion, or the Doppler effect. In the intermediary or kinetic regime, as $y \approx 1$, the analysis is most difficult and the spectral scattering line shape can be derived by solving the Boltzmann equation, for which approximate methods must be employed.

The Tenti-S6 model, proposed in the 1970s~\cite{Boley1972,Tenti1974}, is such a model which has become a standard approach for Rayleigh-Brillouin scattering in the kinetic regime. In this model, the collision integral of Boltzmann equation is replaced by seven or six matrix elements, which can be expressed in terms of the macroscopic transport coefficients, pressure $p$, temperature $T$, shear viscosity $\eta_s$, thermal conductivity $\lambda_{th}$, bulk viscosity $\eta_b$ and the internal molar heat capacity $C_{int}$.

An alternative kinetic approach deals with the Boltzmann equation in replacing the collision operator with a single relaxation-time term \cite{Fernandes2007}. This model builds on the Chapman-Enskog model for solving the Boltzmann equation \cite{Chapman1970} and the Grad'6-moment approximation is employed for calculating the light scattering spectral function \cite{Sugawara1967}.

As a third approach, the model by Hammond and Wiggins~\cite{Hammond1976} based on hydrodynamics involves the vibrational and rotational relaxation, $\tau_{vib}$ and $\tau_{rot}$, as signatures of non-ideal gas effects. The Burnett correction is also added to the Navier-Stokes equation to approximately extend it to the rarified regime~\cite{Desai1972}. This model is valid in the higher pressure regime where the a gaseous medium can be envisioned as a fluid continuum. As was indeed shown \cite{Hammond1976,Wang2017} the Hammond-Wiggins model has an extended application in the realm toward lower pressures, thus forming a bridge between the full hydrodynamic and kinetic regimes.

A rough-sphere model, proposed by Marques~\cite{Marques1999}, was recently applied to describe the RB light scattering spectra in SF$_6$ gas \cite{Wang2017}, in which the molecules exhibit the structure of a regular octahedron with a sulfur atom in the center and six fluorine atoms at vertexes, hence taking the form of a spherical molecule. In such model a dimensionless moment of inertia $\kappa$ is an important and uniquely adjustable parameter.

In the present study the Rayleigh Brillouin scattering profile of nitrous oxide, or dinitrogen monoxide (N$_2$O), is investigated experimentally and results are compared to profiles calculated from the four models mentioned. Here, the N$_2$O molecule is chosen as a scattering species, partly in view of its favorable cross section~\cite{Sneep2005}. Goal is to investigate how the various models can describe the spectral profile for Rayleigh-Brillouin scattering off a linear polyatomic molecule, of a shape strongly deviating from sphericity.

\section{Experiment}
\label{SecN2OblueExperiment}

RB scattering from N$_2$O gas (Praxair, purity 99.7\%) is measured at right angles from a scattering cell with an incident laser beam at $\lambda$ = 403.00 nm. For this, infrared light at 806.00 nm is produced with a Nd:VO$_4$ laser (Millennia-X) pumping a continuous wave Titanium:Sapphire laser (Coherent-699). Its output is converted to the second harmonic in an external frequency-doubling cavity producing power levels of 400-500 mWatt. The spectral bandwidth of the blue laser beam is $\sim 2$ MHz, thus negligible for the analyses in the present study.

The scattering cell, equipped with Brewster-angled windows at entrance and exit ports, is placed inside an enhancement cavity in which the blue light beam is amplified by a factor of ten to reach power levels of 4-5 Watt in the scattering volume. During operation both the frequency-doubling and enhancement cavities are locked by H\"{a}nsch-Couillaud opto-electronics~\cite{Hansch1980}.

Scattered light propagates through a bandpass filter (Materion, T = 90$\%$ at $\lambda$ = 403 nm, bandwidth $\triangle\lambda$ = 1.0 nm) onto a Fabry-Perot interferometer (FPI) via an optical projection system consisting of a number of lenses and pinholes to reduce stray light and contributions from Raman scattering. The geometry is such that the opening angle for in-plane scattering is limited to  0.5 degree to not compromise the instrument resolution. Scattered photons are finally collected on a photomultiplier tube. The FPI is half-confocal and has an effective free spectral range (FSR) of 7.498 GHz, which is determined through scanning the laser frequency over more than 80 modes of the FPI while measuring the laser wavelength by a wavelength meter (ATOS). The instrument width is obtained in two different ways yielding a value of $\sigma_{\nu_{instr}}$ = 126.0 $\pm$ 3.0 MHz (FWHM). First an auxiliary reference beam from the frequency-doubled laser is propagated through the scattering cell at right angles and through the optical system. Subsequently scattered light obtained from a metal needle placed in the scattering center is measured while scanning the laser frequency or the piezo-actuated FPI. This instrument function is verified to exhibit the functional form of an Airy function which may well be approximated by a Lorentzian function.
Further details of the experimental setup and the calibration of the RB-spectrometer haven been described in a technical paper~\cite{Gu2012rsi}.

The scattering angle is an important parameter in RB-scattering since it determines the effective width of the spectral profile through the Doppler condition. From measurements on the geometrical lay-out of the setup, where narrow pinholes determine the beam path, the angle $\theta$ is precalibrated at $\theta = 90 \pm 1^{\circ}$. Calibration measurements on argon gas at 1 bar and comparison with Tenti S6 model calculations were used to assess a precise value for the scattering angle yielding $\theta = 89.6 \pm 0.3^{\circ}$ by fitting to the recorded spectral profile. During the measurements on N$_2$O the angular alignment of the setup was kept fixed.

RB scattering spectral profiles were recorded by piezo-scanning the FPI at integration times of 1 s for each step. Typical detection rates were $\sim 2000$ count/s for conditions of 1 bar pressure. A full spectrum covering 100 consecutive RB-peaks and 10,000 data points was obtained in about 3 h. The piezo-voltage scans were linearized and converted to frequency scale by fitting the RB-peak separations to the calibrated FSR-value. Finally, a collocated spectrum was obtained by cutting and adding all individual recordings over $\sim 100$ RB-peaks~\cite{Gu2012rsi}. This procedure yields a noise level of $\sim 0.4$\% (with respect to peak height) for the 1 bar pressure case. Measurements of the RB-scattering profile were performed for conditions of 0.5-4 bar pressure and room temperatures as listed in Table~\ref{TableN2OblueConditions}.

\begin{table}
{\caption{\label{TableN2OblueConditions} Data sets for RB-scattering measurements in N$_2$O gas recorded under conditions as indicated. $y$ represents the uniformity parameter.}}
\begin{center}
\begin{tabular}{c c c c}
\hline
Data set & $p$(bar) & $T$(K)   & $y$ \\
 \hline
 I  &0.560  & 297.81  & 0.51\\
 II &1.035  & 296.25  & 0.95\\
 III&2.074  & 297.28  & 1.90\\
 IV &3.052  & 297.28  & 2.79\\
 V  &4.194  & 297.28  & 3.84\\
\hline
\end{tabular}
\end{center}
\end{table}

\section{Results}
\label{SecN2OblueResults}

\begin{figure*}
\centering
\includegraphics[scale=0.110]{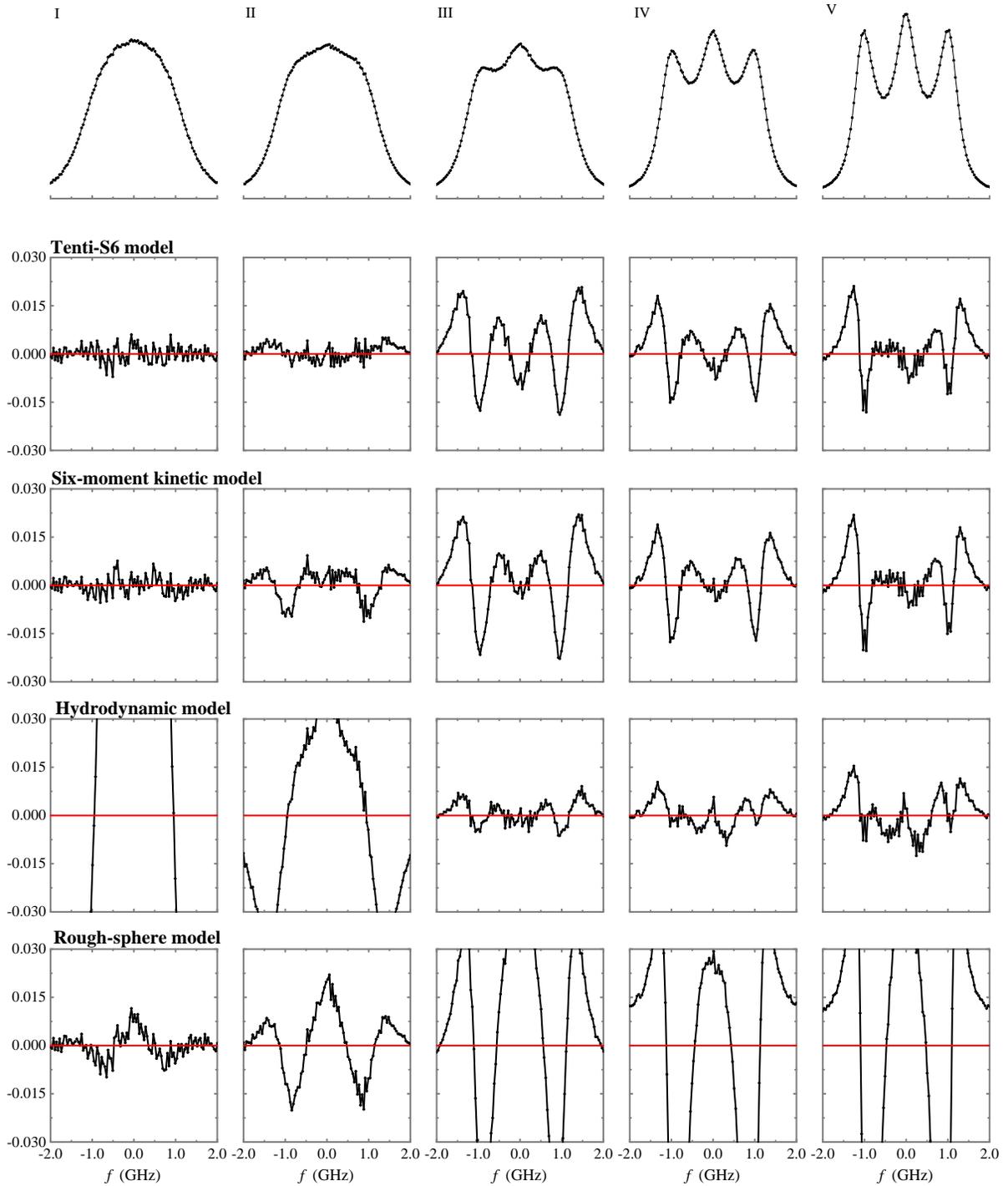}
\caption{Data on RB-scattering in N$_2$O, measured at the five different pressures and temperature conditions (indicated by corresponding Roman numerals I to V as listed in Table~\ref
{TableN2OblueConditions}) at wavelength of $\lambda$ = 403.00 nm and a scattering angle $\theta = 89.6^{\circ}$. Top-row: experimental data on a scale of normalized integrated intensity. Second row: deviations of the Tenti-S6 model description (TS6) as discussed in section \ref{SubSecN2OblueComparewithS6}. Third row: deviations of the six-moment kinetic model as discussed in section \ref{SubSecN2Oblue6-momentkinetic}. Fourth row: deviations from the extended hydrodynamic model (HW) \cite{Hammond1976} as discussed in section \ref{SubSecN2OblueHydrodynamicModel}. Fifth row: deviations from a 'rough spheres model' (RS) as discussed in section~\ref{SecN2OblueRSmodel}. Residuals are plotted on a scale of normalized integrated intensity for each profile.}
\label{FigN2OBlue}
\end{figure*}

The experimental data of the RB-scattering measurements performed on gaseous N$_2$O at five different pressures and all at room temperature are displayed in Fig.~\ref{FigN2OBlue}. The spectral profiles demonstrate the superior quality of the RB-spectrometer in achieving high signal-to-noise reaching a value of $> 100$ even at the lowest pressure (0.5 bar) and better at the higher pressures. While the spectral intensities scale with the familiar factor $(n-1)^2$ in the Rayleigh scattering cross section, with $n$ the index of refraction \cite{Sneep2005}, the profiles are plotted on a normalized scale of unit area. In the following model calculations of the spectral profile are carried out, in the context of (i) the Tenti S6 model~\cite{Boley1972,Tenti1974}, (ii) the Grad's 6-moment kinetic model optimizing a single relaxation time constant~\cite{Fernandes2007}, (iii) a hydrodynamic model as developed in the past for methane gas \cite{Hammond1976}, and (iv) a 'rough-spheres' model as recently developed for poly-atomic molecules \cite{Marques1999} and successfully applied to describe RB scattering in SF$_6$ gas \cite{Wang2017}. Before making the comparison of the model data with the experimental data, the results from the theoretical profiles are folded with a Lorentzian of 126 MHz (FWHM) representing the instrument function of the RB spectrometer.
The comparison is made by calculating the root mean square error (RMSE) expressed as:
\begin{equation}
R_{\rm{rmse}} = \frac{\sqrt{\frac{1}{N}\sum_{i=1}^{N}(I_e(i)-I_m(i))}}{\sum_{i=1}^{N}I_e(i)}
\end{equation}
where $I_{e}$ is the intensity of the experimental spectrum, $I_{m}$ the intensity of modeled spectra and the sum is taken over $N$ data points. While making the comparison between observed and modeled spectra in several cases one or more physical constants (such as the bulk viscosity parameter $\eta_b$) were included in a fitting routine. In the fitting procedures a background zero level was included to account for slight levels of stray light and dark counts on the detector \cite{Gu2012rsi}.

\subsection{The Tenti-S6 model}
\label{SubSecN2OblueComparewithS6}

The Tenti-model was developed in the 1970s to describe the RB-scattering profile of diatomic molecules, in particular for molecular hydrogen~\cite{Boley1972,Tenti1974}. Later this model was revived after investigations were performed involving coherent RB scattering for which RBS codes were developed by Pan et al. \cite{Pan2003,Pan2004}. Those codes were applied to describe both coherent RB scattering \cite{Vieitez2010} and spontaneous RB scattering in CO$_2$ \cite{Gu2014a} and in nitrogen, oxygen and air \cite{Gu2014b}. In those studies it was established that the S6-version of the Tenti model yields a better agreement with experiment than an alternative S7-version.

\begin{table*}
{\caption{\label{TableN2Obulk} Values for the bulk viscosity parameters obtained by fitting the experimental spectra to various model calculations where intramolecular relaxation is included as a free parameter, which is then linked to $\eta_b$. The parameter $\eta_b^T$ is the value obtained by applying the Tenti-S6 model with the bulk viscosity as a single fitting parameter. In the Grad's 6-moment model~\cite{Fernandes2007} the relaxation time $\tau_{int}^{6G}$ is derived as a fitting parameter, and the bulk viscosity $\eta_b^{6G}$ is then derived via Eq.~(\ref{bulk-HW}). In the hydrodynamic model a rotational relaxation time $\tau_{rot}^H$ is derived from a fit to the data and the bulk viscosity $\eta_b^{H}$ is derived via Eq.~(\ref{bulk-HW}). Values in brackets represent the uncertainties derived in a fit. }}
\begin{center}
\begin{tabular}{c c c c c c}
\hline
Data set &  $\eta_b^T$ ($\times 10^{-5}$ Pa$\cdot$s) & $\tau_{int}^{6G}$ ($\times 10^{-10}$ s) & $\eta^{6G}_{b}$ ($\times 10^{-5}$ Pa$\cdot$s)& $\tau_{rot}^H$ ($\times 10^{-10}$ s) & $\eta^{H}_{b}$ ($\times 10^{-5}$ Pa$\cdot$s)\\
 \hline
 I  & 1.06 (0.24)  & 6.95 (0.18) &  0.62 (0.02) \\
 II & 2.63 (0.37)  & 2.49 (0.15) &  0.41 (0.02) \\
 III& 6.18 (2.06)  &             &                & 17.18 (1.39) &  5.70 (0.46) \\
 IV & 5.19 (0.57)  & 15.87 (2.92)  & 7.75 (1.43)  & 13.13 (0.66)  & 6.41 (0.32) \\
 V  & 6.66 (0.56)  & 11.53 (1.24)  & 7.74 (0.83)  & 13.82 (0.80)  & 9.28 (0.54) \\
\hline
\end{tabular}
\end{center}
\end{table*}

The Tenti-model invokes molecular properties in terms of macroscopic transport coefficients and thermodynamic properties, which are usually known or can be measured in a variety of experiments.
The shear viscosity and thermal conductivity of N$_2$O are reported in the literature: $\eta_s = 1.48 \times 10^{-5}$ Pa$\cdot$s and $\lambda_{th} = 17.47 \times 10^{-3}$ W/mK  ~\cite{Uribe1990}.
Alternatively, a value for the thermal conductivity can be estimated by Eucken's formula~\cite{Chapman1970}:
\begin{equation}
\lambda_{th} = \frac{5}{2}\eta_s C_t / M + \rho D(C_{vib}+C_{rot})/M
\label{Eucken}
\end{equation}
where $C_t$, $C_{rot}$, and $C_{vib}$ are the translational, rotational and vibrational molar heat capacities in units of J/mol$\cdot$K, respectively, $M$ is the molar mass ($M=44.01$ g/mol), $\rho$ is mass density and $D$ is the diffusion coefficient~\cite{Uribe1990,Bousheri1987}. This leads to a mass diffusion coefficient of $\rho D$ = 2.09 $\times 10^{-5}$ kg/m$\cdot$s.
If $C_{vib}$ is set to zero in Eq. (\ref{Eucken}) then a value of $\lambda_{th}=14.5 \times 10^{-3}$ W/mK would result, which is in good agreement with the measured value. This illustrates that indeed vibrational relaxation can be ignored as an effective relaxation process.
A measurement of vibrational relaxation yielding $\tau_{vib} = 0.87 \times 10^{-6}$ s~\cite{Dasilva1981} (for a pressure of 1 bar and scaling with $\propto$ 1/$p$) corresponds to a relaxation rate at the MHz level, hence much smaller than the sound frequency, which is at the GHz level. Therefore, here and also below in the other model descriptions, only the rotational relaxation is considered as an internal degree of freedom and the internal molar heat capacity for N$_2$O is $C_{int} = R$, where $R$ is the universal gas constant.

Here we adopt the procedure to regard the final macroscopic transport coefficient, the bulk viscosity $\eta_b$ as a parameter that can be derived via a  least-squares algorithm when comparing the experimental and model spectral profiles. This procedure was followed and documented in studies on the determination of bulk viscosities in N$_2$ \cite{Gu2013b}, in O$_2$ and air~\cite{Gu2013b} and in CO$_2$ \cite{Gu2014a}. For each data set I-V (Table~\ref{TableN2OblueConditions}) $\eta_b$ is  fitted and  results are shown in Table~\ref{TableN2Obulk}.
The deviations between experimental and calculated RB-scattering profiles, for the optimized values of $\eta_b$ for each pressure case, are presented in Fig.~\ref{FigN2OBlue}. For the low pressures of $p=0.5 - 1$ bar near-perfect agreement is found, while for the higher pressures of $p=2-4$ bar still very good agreement is found from the Tenti-S6 model with deviations being smaller than 3\%.

It is found that in this application of the Tenti-S6 model the value of $\eta_b$ appears to be pressure dependent (see Table~\ref{TableN2Obulk}). For the data recorded for $p=0.5$ and $p=1$ bar the $\eta_b$ values are in range $(1-2) \times 10^{-5}$ Pa$\cdot$s, while for the pressures $p=2-4$ bar the $\eta_b$ values level off at $(5-6) \times 10^{-5}$ Pa$\cdot$s. This effect might be considered as a measurement artifact since at the lower pressures collisional relaxation effects are expected to be small and the deviations from a Gaussian profile are also small (see Fig.~\ref{FigN2OBlue}). However, in numerical terms the fits return values  with $<10$\% uncertainties in all cases, so the derivation of bulk viscosities at low pressures should be considered significant.

Interestingly, in the case of CO$_2$ an opposite behaviour was found at low pressures, displaying a decreasing trend in $\eta_b$ with respect to pressure~\cite{Gu2014a}. In the pressure range $p=2-4$ bar a value of $6 \times 10^{-6}$ Pa$\cdot$s was found for CO$_2$, hence an order of magnitude smaller than is presently found for N$_2$O. This is remarkable in view of the similar size and chemical composition of the two molecules.

\subsection{The Grad's six-moment kinetic model}
\label{SubSecN2Oblue6-momentkinetic}

Subsequently we use the kinetic model equation proposed by Fernandes and Marques Jr. \cite{Fernandes2007} to describe the RB scattering profiles measured for N$_2$O. In this model, the collision operator of the Boltzmann equation is replaced by a single relaxation-time term of the form
\begin{equation}
{\mathcal C}(f,f)=-\frac{p}{\eta_s}(f-f_r),
\end{equation}
where $f_r$ is a reference distribution function. The determination of the distribution function follows by requiring that the Chapman-Enskog solution of the model equation to be compatible with Grad's six-moment approximation for polyatomic gases. In the six-moment approximation, the balance equations governing the dynamical behavior of mass density, flow velocity, translational temperature and internal temperature are supplemented with constitutive relations for the pressure tensor, translational heat flux vector and internal heat flux vector.
In their approach, Fernandes and Marques Jr. \cite{Fernandes2007} used Navier-Stokes-Fourier constitutive relations which are valid for the case that the energy exchange between translational and internal degrees of freedom is slow, but not negligible.

The application of Grad's six-moment model equation to describe light scattering experiments in polyatomic gases requires only the
specification of the ratio of heat capacities $\gamma$, the relaxation time of the internal degrees of freedom $\tau_{int}$ and the numerical factor
$u_{\sc 11}^\prime=3\Omega^{\sc (2,2)}/5\Omega^{\sc (1,1)}$ which depends on the law of interaction between molecules through the Chapman-Cowling collision integrals $\Omega^{\sc (\ell,r)}$ \cite{Chapman1970}. For the ratio of heat capacities $\gamma$ (as in the previous model) we take the value 1.4,
while for the numerical factor $u_{\sc 11}^\prime$ we adopt the value 1.32, which follows by assuming that the nitrous oxide molecules interact according to the Lennard-Jones $(6-12)$ model (for details, see Hirschfelder et al. \cite{Hirschfelder1948}).

In the model description the relaxation time $\tau_{\sc \text{int}}$ is then the only unknown variable when comparing with
the experimental data, hence this internal relaxation time is then optimized in a least-squares fitting process. Results of
the fitted values are included in Table \ref{TableN2Obulk} and the calculated RBS profiles are included in Fig.~\ref{FigN2OBlue} in terms of deviations from the experimental spectra. For data set III the model calculation does not converge while performing a least-squares fit to deduce the relaxation parameter. However, in the full range of physically possible relaxation times, hence in range $(3 - 15) \times 10^{-10}$ s, similar $\chi^2$ and $R_{rmse}$ values are determined, and the comparison with the spectrum itself is rather well behaved (see Fig.~\ref{FigN2OBlue}).

The relaxation phenomena in a gas can be described by a single parameter, the bulk viscosity $\eta_b$, which may be related to relaxation times for the degrees of freedom~\cite{Chapman1970}:
\begin{equation}
   \eta_b = \frac{2}{(3+f_{int})^2}p(f_{1}\tau_{1}+f_{2}\tau_{2}+...)
\label{eq:relaxation}
\end{equation}
where $f_{int}$ denotes the internal degrees of freedom of a molecule, $f_{1}$, $f_{2}$, ... are the contributions to $f_{int}$ from the separate parts of the internal energy and $f_{1}+f_{2}+...= f_{int}$. The value 3 in the denominator refers to the three degrees of freedom associated with translational motion and is added to the number of internal degrees of freedom. The values $\tau_{1}$, $\tau_{2}$, ... are the corresponding relaxation times.


In our case we only have rotational degrees of freedom ($f_{rot}=2$ for a linear molecule) as the vibrational modes are frozen. Consequently, Eq. (\ref{eq:relaxation})  reduces to
\begin{equation}
\eta_b^G = \frac{2}{(3+f_{rot})^{2}}pf_{rot}\tau_{rot} = \frac{4}{25} p\tau_{rot}.
\label{bulk-HW}
\end{equation}
Then, the corresponding bulk viscosity $\eta^{G}_{b}$ for data sets III-V can be calculated and values are listed in Table~\ref{TableN2Obulk}.

\subsection{The Hammond-Wiggins hydrodynamic model}
\label{SubSecN2OblueHydrodynamicModel}

In a third approach, the hydrodynamic model of Hammond and Wiggins \cite{Hammond1976} was used to compare with the RB scattering spectral profiles measured for N$_2$O. In this model, vibrational and rotational relaxation times $\tau_{vib}$ and $\tau_{rot}$ are key parameters in the description. Again vibrational relaxation is a too slow process and only rotational relaxation is considered as effectively contributing.

The same code, implemented in a previous analysis on RB-scattering in SF$_6$, was used here to evaluate the 5-component matrix equations of the Hammond-Wiggins model involving fluctuations of the mass density $\bar{\rho}/\rho_0$, translational temperature $\bar{T}/T_0$, momentum or velocity density $\bar{v}/v_0$, vibrational temperature ${\bar T_{\rm{vib}}}/T_0$, and rotational temperature ${\bar T_{\rm{rot}}}/T_0$. Again a value for the thermal conductivity $\lambda_{th}$ was included following Eucken's relation.

In the analysis a fit was made, comparing the experimental and model spectra, in which the rotational relaxation was adopted as a free parameter. The values for $\tau_{rot}$ resulting from the fits are listed in Table~\ref{TableN2Obulk}. At the lowest pressure $p=0.5$ bar the fit did not converge, while for the spectrum recorded at $p=1$ bar an unphysical value was found. Hence these entries are left out of Table \ref{TableN2Obulk}.
In Figure ~\ref{FigN2OBlue} the deviations between experimental and modeled spectra are presented in graphical form. For the data sets III-V, for pressures $p=2-4$ bar, good agreement is found from the HW-model yielding deviations of less than 1\%. This may be considered an excellent result, keeping in mind that an hydrodynamic model focusing on relaxation phenomena is in principle not suited to model dynamics under conditions where collisions play less of a role, like at low pressures. The present study provides a demarcation point, between $p=1$ and $p=2$ bar, where the application of the HW hydrodynamic model becomes relevant. Quantitatively the study provides a value for the rotational relaxation at $\tau \sim 1.5$ ns, which is commensurate with the relaxation found in the Grad's 6-moment model.

Again, from the obtained relaxation times the bulk viscosity $\eta_b^H$ can be derived via Eq. (\ref{eq:relaxation}). Values are listed in Table \ref{TableN2Obulk}.


\subsection{The 'rough-sphere' model}
\label{SecN2OblueRSmodel}

\begin{figure}
\centering
\includegraphics[scale=0.26]{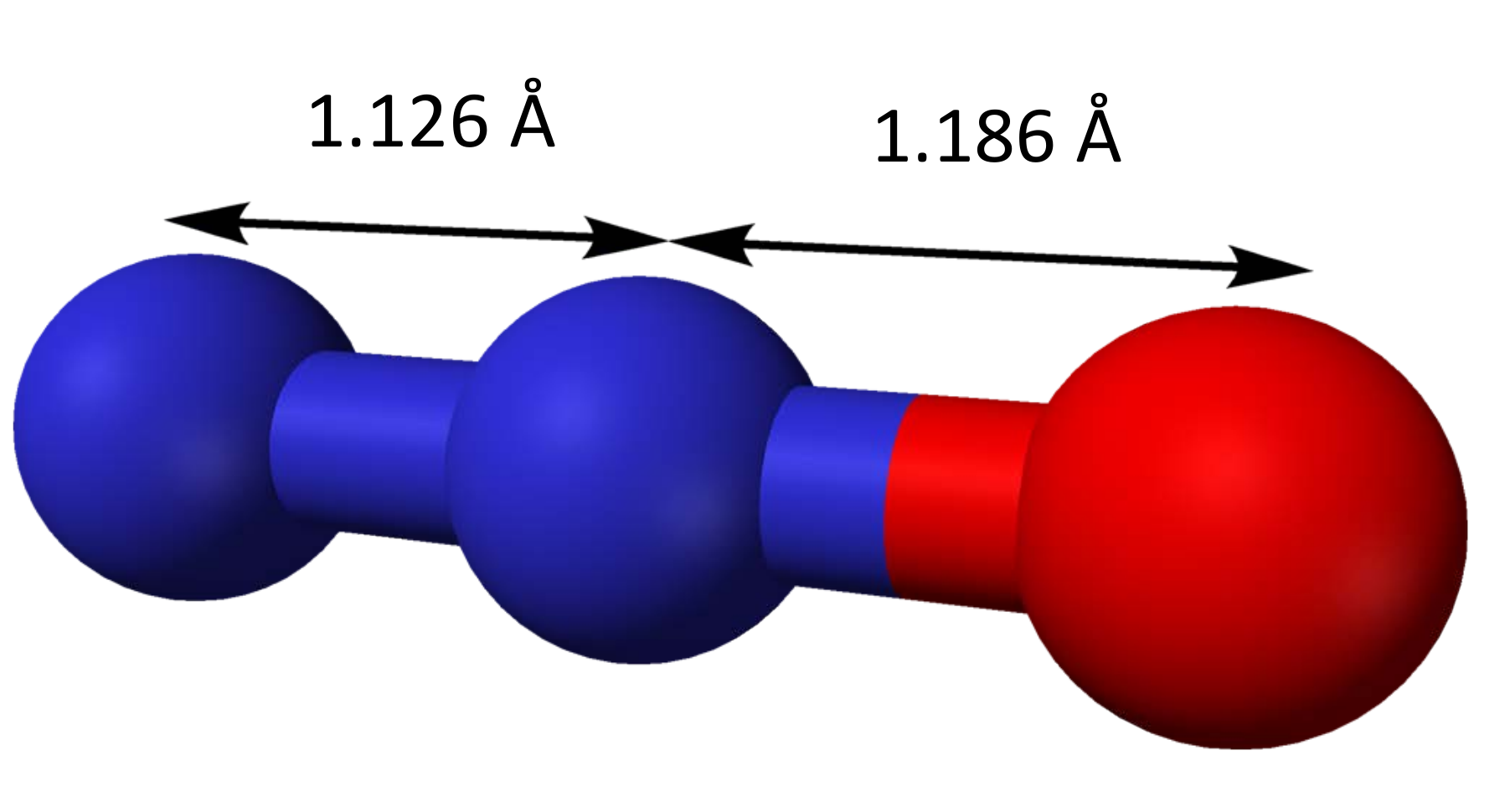}
\caption{Geometrical structure of the linear N$_2$O molecule (in its electronic ground state) with nitrogen atoms depicted as the smaller (blue) balls and the oxygen atom as the larger (red) ball. Internuclear separations for the ground state are displayed as resulting from spectroscopy \cite{Herzberg1966,NIST}.}
\label{FigN2OGeo}
\end{figure}

The rough-sphere model proposed by Marques~\cite{Marques1999}, is also a kinetic theory describing the density fluctuation in polyatomic gases. In this model, a simple relaxation term $\delta(f-f_{\rm{r}})$ is used to replace the collision operator in Boltzmann equation with $f(\vec{r},\vec{v})$ the six-dimensional position-velocity distribution and  $f_{\rm{r}}$ representing a reference distribution function. Here the coefficients satisfy the conservation laws, while the collisional transfer of momentum and energy agree with the full Boltzmann description. The rough sphere model considers the interaction between the translational and rotational degrees of freedom and regards the collisions between molecules as hard spheres, thereby ignoring the effect of vibrational relaxation. This model is built on a dimensionless moment of inertia $\kappa=4I/md^2$, with $m$ the mass, $I$ the moment of inertia and $d$ an effective diameter of the molecule. The moment of inertia $I$ may be derived from the rotational constant as obtained in microwave spectroscopy of the molecule for which a value of $B=12561$ MHz was reported \cite{Andreev1976}, corresponding to a moment of inertia of $I = 66.7\times 10^{-47}$ kg$\cdot$m$^{2}$ for N$_2$O \cite{Herzberg1966,NIST}.
In Fig.~\ref{FigN2OGeo} the geometrical structure of the N$_2$O molecule is depicted with internuclear separations between nitrogen and oxygen atoms. The distance between outer atoms is $2.3$ \AA, but the effective diameter $d$ of the molecule is determined in a number of studies to be higher: $d=3.85$ {\AA}~\cite{Baalbaki1986} and $d= 3.828$ \AA~\cite{Reid1987}. This results in a value of $\kappa= 0.246$.

With a value for the shear viscosity of $\eta_s=1.48 \times 10^{-5}$ Pa$\cdot$s the rough sphere model derives a value for internal relaxation effects, represented as a bulk viscosity via the relation \cite{Chapman1970}
\begin{equation}
\eta_b = \eta_s \frac{6+13\kappa}{60\kappa}
\end{equation}
resulting in a value of $\eta_b=0.92 \times 10^{-5}$ Pa$\cdot$s.
Model spectra for RB-scattering in N$_2$O were subsequently calculated using the formalism presented by Marques \cite{Marques1999,Marques1993}.
Results are displayed in Fig.~\ref{FigN2OBlue} in terms of deviations between experimental and modeled spectral profiles. Large discrepancies arise in particular at the high pressure values where relaxation phenomena play an important role. Numerically the rough sphere model delivers a value for the bulk viscosity $\eta_b$ that is significantly smaller than the values obtained from the kinetic Tenti-S6 model, by a factor of six. Hence relaxation phenomena may be underestimated in the description.
The relaxation phenomena are not well described by the assumptions made for collisions to occur between object of spherical nature. In view of the geometrical structure of the N$_2$O molecule as displayed in Fig.~\ref{FigN2OGeo} this is not surprising.

Lastly, it is mentioned that the discrepancy resulting from the rough sphere model is not just based on the non-sphericity of the N$_2$O geometrical structure. The model predicts a value for the heat capacity ratios of $\gamma = 4/3$, which is slightly smaller than the
value $\gamma = 7/5$ for nitrous oxide. Therefore, the rough sphere model predicts a wrong value for the speed of sound and this is crucial in producing good agreement with experiments.

\section{Discussion and Conclusion}
\label{SecN2OblueDiscussandConclusion}

In this study spontaneous Rayleigh-Brillouin scattering spectra of N$_2$O gas of high signal-to-noise are experimentally recorded, allowing for detailed comparison with models describing the phenomena underlying the scattering profiles. RB-scattering is a complex phenomenon as it entails all intramolecular and intermolecular interactions of molecules in a dense gaseous environment. Hence, the scattering profiles involve information on the spectroscopy, internal vibrational and rotational relaxation, coupling with translational motion, quantum state-to-state energy transfer, velocity-changing collisions, etcetera. The entirety of the behavior can in principle be described by the Boltzmann equation. Since the full six-dimensional information on position-velocity coordinates and involving all state-to-state collisional cross sections in a medium are not known, approximate methods must be invoked to model the RB light scattering process.

In the present study four of such prevailing models are applied to the scattering of the nitrous oxide molecule which is special for a number of reasons. N$_2$O is a polyatomic molecule for which the linear N-N-O structure (see also Fig.~\ref{FigN2OGeo}) does not provide a symmetry point as is the case for CO$_2$ \cite{Gu2014a}. While a number of recent studies were performed to model RB-scattering in diatomic molecules \cite{Boley1972,Tenti1974,Sandoval1976,Gu2014b} the quest is now to investigate RB-scattering in polyatomic molecules of different symmetry and sizes. N$_2$O is a convenient target in view of its large scattering cross section \cite{Sneep2005}.

\begin{figure}
\centering
\includegraphics[scale=0.32]{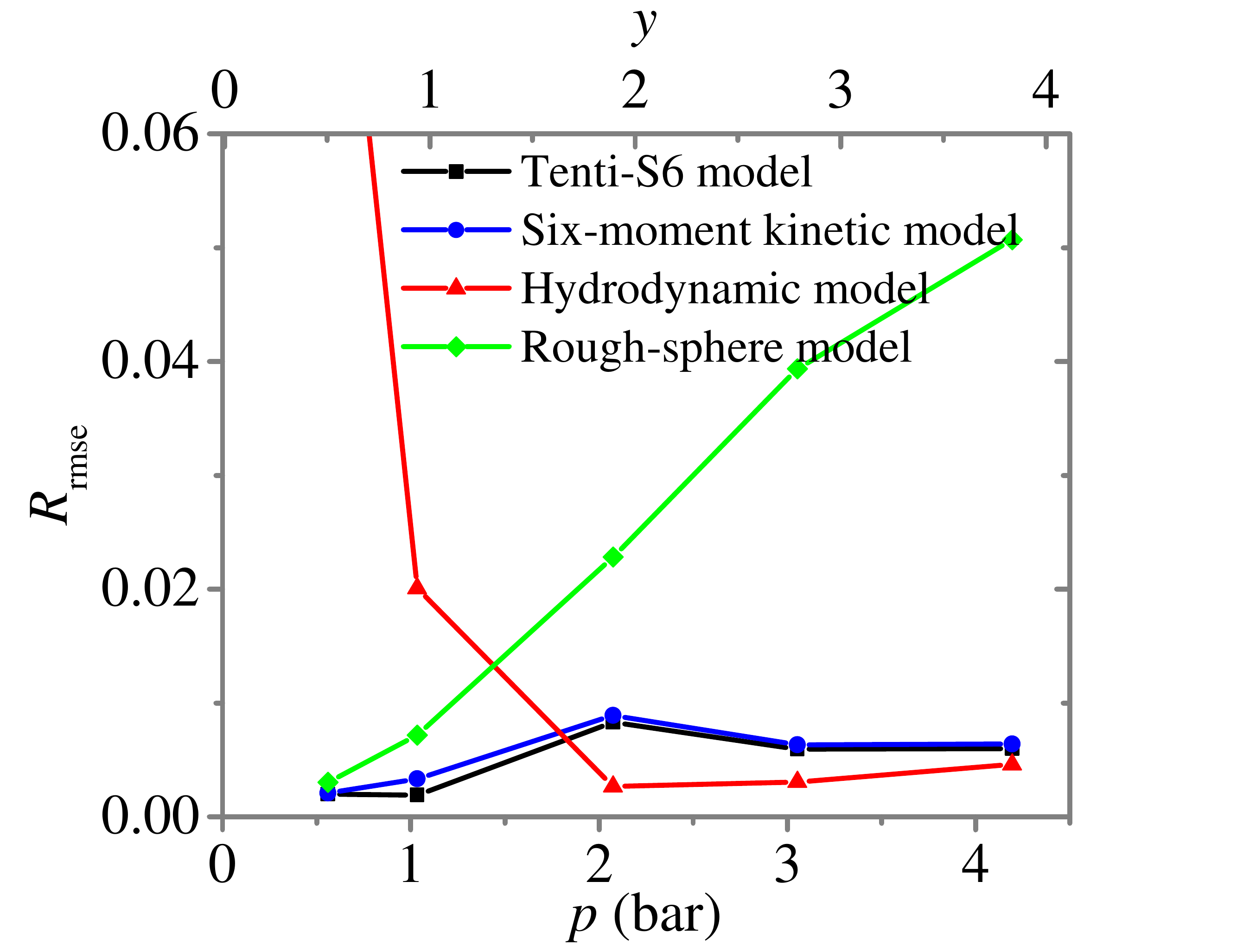}
\caption{The root-mean-square error based on the deviation of experimental spectra and the models (Tenti-S6, the Grad's 6-moment kinetic model, the HW-hydrodynamic and the rough sphere models) after folding with the instrument function. The top-axis shows the scale converted to the uniformity parameter $y$.}
\label{FigN2OBlueRMSE}
\end{figure}

The deviations between the experimental spectra and the modeled spectra, for the models discussed in the present study, are presented in condensed form in terms of a normalized root-mean-square error $R_{\rm{rmse}}$ value of each model at the five pressure conditions is shown in value Fig.~\ref{FigN2OBlueRMSE}.
The Tenti-S6 model provides overall a good description of RB-scattering, now also for the polyatomic N$_2$O molecule. For the lowest pressures, where the spectral profile is close to Gaussian, $R_{\rm{rmse}}$ deviations are below the 1\% level. For the higher pressure ranges deviations grow to the 2\% level, but in view of the quality of the spectra, deviations are significant. The value for the bulk viscosity $\eta_b$ is found to be pressure dependent, as expected from general considerations on relaxation phenomena \cite{Chapman1970}, while leveling off to a value of $6 \times 10^{-5}$ Pa$\cdot$s. This is an order of magnitude larger than the bulk viscosity of CO$_2$, a molecule of comparable size and composition.

The Grad's six-moment kinetic model shows a similarly good performance in the comparison with experimental data as does the Tenti-S6 model. In fact the spectral comparisons in Fig.~\ref{FigN2OBlue} as well as the values for the root-mean-square deviations, as in Fig.~\ref{FigN2OBlueRMSE} are virtually identical. As for the underlying physical parameter, the bulk viscosity as deduced from the internal relaxation number $Z$, some difference is found with the values derived in the Tenti-S6 model. However this is mainly the case for the low pressures $p=0.5 - 1$ bar, where collisional relaxation does not play a decisive role, and the value of $\eta_b$ barely affects the spectra profile. For the higher pressures, $p=3-4$ bar, where collisional relaxation is more decisive the differences between bulk viscosities $\eta_b^{T}$ and $\eta_b^{6G}$ are some 20-30\%.

The Hammond-Wiggins hydrodynamic model is well applicable to the spectra measured at pressures at 2-4 bar, where agreement is found below the 1\% level. In fact, at pressures $p=2-4$ bar this model yields the best description of RB light scattering in N$_2$O gas. This hydrodynamic model is not applicable at the low pressures $p \leq 1$ bar, where extremely large deviations are found between modeled and experimental spectra, even when a rotational relaxation parameter is adapted in a fit.
However, the non-applicability of a hydrodynamic model in a low-pressure regime is well understandable.
Also in the context of this hydrodynamic model the relaxation phenomenon can be connected to a bulk viscosity parameter, values for which are listed in Table~\ref{TableN2Obulk}.

Although the three models, Tenti-S6, Grad's 6 moment and HW-hydrodynamic, have very different physical basis and show differing deviations between model and experimental data, in the range $p=2-4$ bar the internal relaxation is described by a gas transport coefficient that is bounded within some limits. The value for the bulk viscosity pertaining to all three descriptions is bound by $\eta_b \sim (6 \pm 2) \times 10^{-5}$ Pa$\cdot$s.

The rough-sphere model turns out to be not applicable to describe RB light scattering in N$_2$O gas. Even in the near-collisionless regime of $p=0.5$ bar deviations are evident, but they grow to large proportions for increasing pressures. This is, after all, not surprising, in view of the non-spherical geometrical structure of the N$_2$O molecule.

\section*{Supplementary Material}

The experimental Rayleigh-Brillouin scattering data as displayed in the top column of Fig.~\ref{FigN2OBlue} measured for the five pressures are provided as Supplementary Material to this article.

\section*{Acknowledgements}

This research was supported by the China Exchange Program jointly run by the Netherlands Royal Academy of Sciences (KNAW) and the Chinese Ministry of Education. YW acknowledges support from the Chinese Scholarship Council (CSC) for his stay at VU Amsterdam.
WU acknowledges the European Research Council for an ERC-Advanced grant under the European
Union's Horizon 2020 research and innovation programme (grant agreement No 670168).

\section*{References}


\end{document}